\documentclass[nolinenumbers,trackchanges,twocolumn]{aastex701}

\usepackage[utf8]{inputenc}
\usepackage{graphicx}
\graphicspath{{figures/}{./}{../figures/}}
\usepackage{tabularx}
\usepackage{newtxtext}
\usepackage[T1]{fontenc}

\usepackage{color}
\usepackage[shortlabels]{enumitem}
\usepackage{subfig} % Added by Ramon for IRIS part
\usepackage{datetime}

\usepackage{pifont}% http://ctan.org/pkg/pifont
\usepackage{siunitx}

\usepackage[nolist]{acronym}

\definecolor{mpg}{rgb}{0.,0.48627451,0.4745098}
\definecolor{mpgdark}{rgb}{0.,0.36470588, 0.35588235}
\definecolor{linka}{rgb}{0.0,0.0,0.7}
\definecolor{linkb}{rgb}{0.0,0.0,0.5}
\definecolor{linkc}{rgb}{0.0,0.0,0.3}
\definecolor{cmt}{rgb}{0.5,0.0,0.0}
\definecolor{todo}{rgb}{0.6,0.4,0.8}
\definecolor{shade}{rgb}{0.7,0.7,0.7}
\definecolor{akl}{rgb}{0.6,0.2,0.0}
\definecolor{jcdti}{rgb}{0.6,0.2,0.6}
\definecolor{sn}{rgb}{0.8,0.2,0.6}
\definecolor{sks}{rgb}{0.8,0.1,0.0}
\definecolor{pnb}{rgb}{1.0,0.49,0.0}

\newcommand{\colfig}[3][1.0]{
  \begin{figure}
    \centering
    \includegraphics[width=#1\linewidth,clip=TRUE]{#2}
    \caption{#3}
    \label{#2}
  \end{figure}
}
\newcommand{\colfigwide}[3][1.0]{
  \begin{figure*}
    \centering
    \includegraphics[width=#1\linewidth,clip=TRUE]{#2}
    \caption{#3}
    \label{#2}
  \end{figure*}
}
\newcommand{\colfigtwo}[4][1.0]{
  \begin{figure}
    \centering
    \includegraphics[width=#1\linewidth,clip=TRUE]{#2}
    \includegraphics[width=#1\linewidth,clip=TRUE]{#3}
    \caption{#4}
    \label{#2}
  \end{figure}
}

\newcommand{\change}[1]{\textcolor{mpg}{\textit{#1}}}

\newcommand{\sks}[1]{{\color{red}\textit{Sami: #1}}}
\newcommand{\akl}[1]{{\color{akl}\textit{Andi: #1}}}
\newcommand{\al}[1]{{\color{akl}\textit{Andi: #1}}}
\newcommand{\dpr}[1]{{\color{mpg}\textit{#1}}}
\newcommand{\RHC}[1]{{\color{mpg}\textit{#1}}}

\renewcommand{\sks}[1]{}
\renewcommand{\akl}[1]{}
\renewcommand{\al}[1]{}
\renewcommand{\dpr}[1]{}
\renewcommand{\RHC}[1]{}
\renewcommand{\change}[1]{#1}

\newcommand{\figref}[1]{Fig.~\ref{#1}} %Reference for figure
\newcommand{\tabref}[1]{Tab.~\ref{#1}} %Reference for figure
\newcommand{\sref}[1]{Sect.~\ref{#1}} %Reference for figure
\newcommand{\sunrise}{\textsc{Sunrise}}

\newcommand{\sunriseii}{\textsc{Sunrise~ii}}
\newcommand{\sunriseiii}{\textsc{Sunrise~iii}}
\newcommand{\hinode}{\textit{Hinode}}
\newcommand{\halpha}{H$\alpha$}

\newcommand{\kms}{km\,s$^{-1}$}

\renewcommand{\ion}[2]{#1\,\textsc{#2}}

\newcommand{\caiih}{Ca\,\textsc{II}\,H}

\newcommand{\fei}{Fe\,\textsc{I}}

\begin{acronym}
\acro{aia}[AIA]{Atmospheric Imaging Assembly}
\acro{alma}[ALMA]{Atacama Large Millimeter/sub-millimeter Array}
\acro{amhd}[AMHD]{{Analog, Motor, and Heaters Drivers}}
\acro{ao}[AO]{adaptive optics}
\acro{bbso}[BBSO]{Big Bear Solar Observatory} 
\acro{bpo}[BPO]{Balloon Program Office}
\acro{cad}[CAD]{Computer Aided Design}
\acro{cfrp}[CFRP]{carbon fiber-reinforced polymers}
\acro{chase}[CHASE]{Chinese \halpha{} Solar Explorer}
\acro{cmos}[CMOS]{complementary metal oxide semiconductor}
\acro{cnoc}[CNOC]{cold non-operational case}
\acro{coc}[COC]{cold operational case}
\acro{coi}[CoI]{Co-Investigator}
\acro{csbf}[CSBF]{Columbia Scientific Ballooning Facility}
\acro{ct}[CT]{correlation tracker}
\acro{cws}[CWS]{Correlating Wavefront Sensor}
\acro{dkist}[DKIST]{Daniel K. Inouye Solar Telescope}
\acro{docdb}[S3-DocDB]{\sunriseiii{} Document Database}
\acro{dss}[DSS]{Data Storage System}
\acro{dst}[DST]{Dunn Solar Telescope}
\acro{egse}[EGSE]{electrical ground support equipment}
\acro{elink}[E-Link]{Esrange Airborne Data Link}
\acro{eoc}[EOC]{Esrange Operations Center}
\acro{erack}[E-rack]{electronic rack}
\acro{esrange}[Esrange]{Esrange Space Center}
\acro{evtm}[EVTM]{Ethernet Via Telemetry}
\acro{falc}[FAL-C]{FAL-C}
\acro{fem}[FEM]{finite element model}
\acro{fft}[FFT]{fast Fourier transform}
\acro{fits}[FITS]{Flexible Image Transport System}
\acro{fov}[FoV]{field-of-view}
\acro{fps}[fps]{frames per second}
\acro{fram}[FRAM]{ferroelectric random access memory}
\acro{frr}[FRR]{Flight Readiness Review}
\acro{fsm}[FSM]{finite state machine}
\acro{fts}[FTS]{Fourier Transform Spectrograph}
\acro{fwhm}[FWHM]{full width at half maximum}
\acro{goc}[GOC]{Göttingen Operations Center}
\acro{gong}[GONG]{Global Oscillation Network Group}
\acro{gps}[GPS]{Global Positioning System}
\acro{gregor}[GREGOR]{GREGOR}
\acro{gst}[GST]{Goode Solar Telescope}
\acro{gusto}[GUSTO]{Galactic/Extragalactic ULDB Spectroscopic Terahertz Observatory}
\acro{hmi}[HMI]{Helioseismic Magnetic Imager}
\acro{hnoc}[HNOC]{hot non-operational case}
\acro{hoc}[HOC]{hot operational case}
\acro{hrw}[HRW]{Heat Rejection Wedge}
\acro{hvps}[HVPS]{high voltage power supply}
\acro{iaa}[IAA]{Instituto de Astrofísica de Andalucía}
\acro{iac}[IAC]{Instituto de Astrofísica de Canarias}
\acro{ibis}[IBIS]{Interferometric BIdimensional Spectropolarimeter}
\acro{ics}[ICS]{Instrument Control System}
\acro{icu}[ICU]{Instrument Control Unit}
\acro{ihop}[IHOP]{IRIS/\hinode{} Operations Plan}
\acro{imax}[\textsc{IMaX}]{Imaging Magnetograph eXperiment}
\acro{imtek}[IMTEK]{Department of Microsystems Engineering}
\acro{imu}[IMU]{Inertial Measurement Unit}
\acro{inta}[INTA]{Instituto Nacional de Técnica Aeroespacial}
\acro{iridium}[Iridium Pilot]{Iridium Pilot\textsuperscript{\textregistered}}
\acro{iris1}[IRIS-1]{Image Recording Instrument for \sunriseii{}}
\acro{iris2}[IRIS-2]{Image Recording Instrument for \sunriseiii{}}
\acro{iris}[IRIS]{Interface Region Imaging Spectrograph}
\acro{ir}[IR]{infrared}
\acro{islid}[\textsc{ISLiD}]{Image Stabilization and Light Distribution unit}
\acro{jaxa}[JAXA]{Japan Aerospace Exploration Agency}
\acro{apl}[APL]{Johns Hopkins University Applied Physics Laboratory}
\acro{kbsi}[KBSI]{Korea Basic Science Institute}
\acro{kis}[KIS]{Institut für Sonnenphysik}
\acro{lcvr}[LCVR]{liquid crystal variable retarder}
\acro{ldb}[LDB]{long-distance balloon}
\acro{ld}[LD]{launch date}
\acro{led}[LED]{light-emitting diode}
\acro{lldpe}[LLDPE]{linear low-density polyethylene}
\acro{los}[LoS]{line-of-sight}
\acro{lte}[LTE]{local thermodynamic equilibrium}
\acro{mast}[MAST]{Multi-Application Solar Telescope}
\acro{mdi}[MDI]{Michelson Doppler Imager}
\acro{mhd}[MHD]{magneto-hydrodynamic}
\acro{MHD}[MHD]{Magneto-Hydrodynamic}
\acro{mit}[MIT]{Massachusetts Institute of Technology}
\acro{mli}[MLI]{multilayer insulation}
\acro{momfbd}[MOMFBD]{Multi-Object Multi-Frame Blind Deconvolution}
\acro{mpae}[MPAe]{Max-Planck-Institut für Aeronomie}
\acro{mpg}[MPG]{Max-Planck-Gesellschaft}
\acro{mps}[MPS]{Max Planck Institute for Solar System Research}
\acro{mrr}[MRR]{Mission Readiness Review}
\acro{mtc}[MTC]{Main Telescope Controller}
\acro{mtu}[MTU]{Momentum Transfer Unit}
\acro{muramce}[MURaM-ChE]{MURaM Chromospheric Extension}
\acro{muram}[MURaM]{MPS/University of Chicago Radiative MHD}
\acro{naoj}[NAOJ]{National Astronomical Observatory of Japan}
\acro{nasa}[NASA]{National Aeronautics and Space Administration}
\acro{nic}[NIC]{network interface controller}
\acro{nir}[near-IR]{near-infrared}
\acro{nlte}[non-LTE]{non-local thermodynamic equilibrium}
\acro{nuv}[near-UV]{near-ultraviolet}
\acro{nvst}[NVST]{New Vacuum Solar Telescope}
\acro{occ}[OCC]{Operations Control Center}
\acro{olr}[OLR]{outgoing longwave radiation}
\acro{pbs}[PBS]{polarising beam-splitter}
\acro{pcb}[PCB]{printed circuit board}
\acro{pcu}[PCU]{Power Converter Unit}
\acro{pdu}[PDU]{Power Distribution Unit}
\acro{pfi}[PFI]{Post Focus Instrumentation}
\acro{phi}[PHI]{Polarimetric and Helioseismic Imager}
\acro{pi}[PI]{Principal Investigator}
\acro{pmu}[PMU]{Polarization Modulation Unit}
\acro{poe}[PoE]{Power over Ethernet}
\acro{ppd}[PPD]{PFI Power Distribution Unit}
\acro{psg}[PSG]{Polarization State Generator}
\acro{psf}[PSF]{point spread function}
\acro{pcs}[PCS]{Pointing Control System}
\acro{hk}[HK]{housekeeping}
\acro{ramon}[RAMON]{RAdiation MONitor}
\acro{rms}[rms]{root-mean-square}
\acro{rtc}[RTC]{real-time clock}
\acro{s2n}[S/N]{signal-to-noise}
\acro{s3pc}[S$^3$PC]{Spanish Space Solar Physics Consortium}
\acro{scip}[\textsc{SCIP}]{\sunrise{} Chromospheric Infrared spectro-Polarimeter}
\acro{sdo}[SDO]{Solar Dynamics Observatory}
\acro{sgt}[SGT]{Sun Guider Telescope}
\acro{sjc}[SJC]{Slit-Jaw Camera}
\acro{sipm}[SiPM]{Silicon PhotoMultiplier}
\acro{sip}[SIP]{Support Instrumentation Package}
\acro{smart}[SMART]{Solar Magnetic Activity Research Telescope}
\acro{smm}[SMM]{scan mirror mechanism}
\acro{soho}[SOHO]{Solar and Heliospheric Observatory}
\acro{solo}[Solar Orbiter]{Solar Orbiter}
\acro{sophi}[SO/PHI]{Solar Orbiter Polarimetric and Helioseismic Imager}
\acro{sotnfi}[NFI]{Narrowband Filter Imager}
\acro{sotsp}[SP]{Spectro-Polarimeter}
\acro{sot}[SOT]{Solar Optical Telescope}
\acro{spg}[SPG]{Solar Physics Group}
\acro{spgcam}[SPGCam]{Solar Physics Group Camera}
\acro{ssct}[SSCT]{Science/Support Compatibility Test}    
\acro{ssc}[SSC]{Swedish Space Corporation}
\acro{ssd}[SSD]{solid state disk}
\acro{ssh}[SSH]{Secure Shell}
\acro{ssm}[SSM]{second surface mirror}
\acro{sst}[SST]{Swedish Solar Telescope}
\acro{sswg}[SSWG]{\sunrise{} Science Working Group}
\acro{ss}[SS]{Science Stack}
\acro{starlink}[Starlink]{Starlink (a division of SpaceX)}
\acro{sufi}[\textsc{SuFI}]{\sunrise{} Filter Imager}
\acro{sppt}[\textsc{SPPT}]{\sunriseiii{} Science Planning and Pointing Tool}
\acro{sunrise}[\textsc{Sunrise}]{{\textbf{S}olar \textbf{U}V to \textbf{N}ear-infra\textbf{R}ed \textbf{I}maging and \textbf{S}pectropolarimetric \textbf{E}xploration}}
\acro{susi}[\textsc{SUSI}]{\sunrise{} \acs{uv} Spectropolarimeter and Imager}
\acro{suit}[SUIT]{Solar Ultraviolet Imaging Telescope}
\acro{suvi}[SUVI]{Solar Ultraviolet Imager}
\acro{sutri}[SUTRI]{Solar Upper Transition Region Imager}
\acro{tac}[TAC]{Telescope Allocation Committee}
\acro{tc}[TC]{Telecommand}
\acro{tdrss}[TDRSS]{Tracking \& Data Relay Satellite System}
\acro{tmtc}[TM/TC]{Telemetry and Telecommand}
\acro{tm}[TM]{Telemetry}
\acro{tumag}[\textsc{TuMag}]{Tunable Magnetograph}
\acro{tv}[TV]{thermal-vacuum}
\acro{unival}[UV]{Universitat de València}
\acro{upm}[UPM]{Universidad Politécnica de Madrid}
\acro{usaf}[USAF]{U.S. Air Force}
\acro{uv}[UV]{ultraviolet}
\acro{vda}[VDA]{vapor deposited aluminum}
\acro{vpn}[VPN]{virtual private network}
\acro{vtt}[VTT]{German Vacuum Tower Telescope} 
\acro{wfs}[WFS]{wavefront sensor}

\end{acronym}
\newcommand{\susi}{\ac{susi}}

\newcommand{\npeaks}{502}
\newcommand{\tres}{0.51}
\newcommand{\nsteps}{6981}
\newcommand{\nlines}{19}
\newcommand{\ncatone}{5.1\%}
\newcommand{\ncattwo}{15.0\%}
\newcommand{\ncatthree}{79.9\%}
\newcommand{\vbingreen}{80\%}
\newcommand{\vbinyellow}{10\%}
\newcommand{\vbinpurple}{10\%}
\newcommand{\vhigh}{0.035}
\newcommand{\vlow}{0.01}
\newcommand{\tlagthresh}{6}

\begin{document}

\title{Height Dependent Phase Shifts of Wave Pulses in the Lower Solar Atmosphere Measured with \sunriseiii{}}

\correspondingauthor{Andreas~Lagg}
\author[orcid=0000-0003-1459-7074,sname='Lagg']{Andreas~Lagg} \affiliation{Max-Planck-Institut für Sonnensystemforschung, Justus-von-Liebig-Weg 3, 37077 Göttingen, Germany}\email[show]{lagg@mps.mpg.de}

%%% main contributors
\author[orcid=0000-0003-3490-6532,sname='Smitha']{H.~N.~Smitha} \affiliation{Max-Planck-Institut für Sonnensystemforschung, Justus-von-Liebig-Weg 3, 37077 Göttingen, Germany}\email{narayanamurthy@mps.mpg.de}

%%% Principal Investigator:
\author[orcid=0000-0002-3418-8449,sname='Solanki']{Sami~K.~Solanki} \affiliation{Max-Planck-Institut für Sonnensystemforschung, Justus-von-Liebig-Weg 3, 37077 Göttingen, Germany}\email{solanki@mps.mpg.de}

%%% Lead-CoIs
\author[orcid=0000-0001-6317-4380,sname='Riethmüller']{Tino~L.~Riethmüller} \affiliation{Max-Planck-Institut für Sonnensystemforschung, Justus-von-Liebig-Weg 3, 37077 Göttingen, Germany}\email{riethmueller@mps.mpg.de}
\author[orcid=0000-0002-9972-9840,sname='Gandorfer']{Achim~Gandorfer} \affiliation{Max-Planck-Institut für Sonnensystemforschung, Justus-von-Liebig-Weg 3, 37077 Göttingen, Germany}\email{gandorfer@mps.mpg.de}
\author[orcid=0009-0009-4425-599X,sname='Feller']{Alex~Feller} \affiliation{Max-Planck-Institut für Sonnensystemforschung, Justus-von-Liebig-Weg 3, 37077 Göttingen, Germany}\email{feller@mps.mpg.de}
\author[orcid=0000-0003-1409-1145,sname='Iglesias']{Francisco~A.~Iglesias} \affiliation{Max-Planck-Institut für Sonnensystemforschung, Justus-von-Liebig-Weg 3, 37077 Göttingen, Germany}\affiliation{Grupo de Estudios en Heliofísica de Mendoza, CONICET, Universidad de Mendoza, Boulogne sur Mer 683, 5500 Mendoza, Argentina}\email{iglesias@mps.mpg.de}
\author[orcid=0000-0003-0175-6232,sname='Siu-Tapia']{Azaymi~L.~Siu-Tapia} \affiliation{Instituto de Astrofísica de Andalucía, CSIC, Glorieta de la Astronomía s/n, 18008 Granada, Spain}\affiliation{Spanish Space Solar Physics Consortium}\email{siu@iaa.es}
\author[orcid=0000-0002-3387-026X,sname='del~Toro~Iniesta']{Jose~Carlos~del~Toro~Iniesta} \affiliation{Instituto de Astrofísica de Andalucía, CSIC, Glorieta de la Astronomía s/n, 18008 Granada, Spain}\affiliation{Spanish Space Solar Physics Consortium}\email{jti@iaa.es}
\author[orcid=0000-0002-5054-8782,sname='Katsukawa']{Yukio~Katsukawa} \affiliation{National Astronomical Observatory of Japan, 2-21-1 Osawa, Mitaka, Tokyo 181-8588, Japan}\affiliation{Department of Astronomy, The University of Tokyo, 7-3-1, Hongo, Bunkyo-ku, Tokyo 113-0033, Japan}\affiliation{Department of Astronomical Science, The Graduate University for Advanced Studies (SOKENDAI), 2-21-1 Osawa, Mitaka, Tokyo 181-8588, Japan}\email{yukio.katsukawa@nao.ac.jp}
\author[orcid=0000-0002-0787-8954,sname='Bernasconi']{Pietro~Bernasconi} \affiliation{Johns Hopkins University Applied Physics Laboratory, 11100 Johns Hopkins Road, Laurel, Maryland, USA}\email{pietro.bernasconi@jhuapl.edu}
\author[sname='Berkefeld']{Thomas~Berkefeld} \affiliation{Institut für Sonnenphysik (KIS), Georges-Köhler-Allee 401a, 79110 Freiburg, Germany}\email{thomas.berkefeld@leibniz-kis.de}
\author[orcid=0000-0001-5616-2808,sname='Kubo']{Masahito~Kubo} \affiliation{National Astronomical Observatory of Japan, 2-21-1 Osawa, Mitaka, Tokyo 181-8588, Japan}\affiliation{Department of Astronomical Science, The Graduate University for Advanced Studies (SOKENDAI), 2-21-1 Osawa, Mitaka, Tokyo 181-8588, Japan}\email{masahito.kubo@nao.ac.jp}
\author[orcid=0000-0001-8829-1938,sname='Orozco~Suárez']{David~Orozco~Suárez} \affiliation{Instituto de Astrofísica de Andalucía, CSIC, Glorieta de la Astronomía s/n, 18008 Granada, Spain}\affiliation{Spanish Space Solar Physics Consortium}\email{orozco@iaa.es}

% MPS Helpers outside Sunrise

\author[orcid=0000-0001-9474-8447,sname='Cameron']{Robert~Cameron} \affiliation{Max-Planck-Institut für Sonnensystemforschung, Justus-von-Liebig-Weg 3, 37077 Göttingen, Germany}\email{cameron@mps.mpg.de}

\author[orcid=0000-0002-2391-6156,sname='Schou']{Jesper~Schou} \affiliation{Max-Planck-Institut für Sonnensystemforschung, Justus-von-Liebig-Weg 3, 37077 Göttingen, Germany}\email{schou@mps.mpg.de}

\author[orcid=0000-0003-1670-5913,sname='Przybylski']{Damien~Przybylski} \affiliation{Max-Planck-Institut für Sonnensystemforschung, Justus-von-Liebig-Weg 3, 37077 Göttingen, Germany}\email{przybylski@mps.mpg.de}

%%% Sunrise CoIs
\author[orcid=0000-0001-9228-3412,sname='Álvarez-Herrero']{Alberto~Álvarez-Herrero} \affiliation{Instituto Nacional de T\'ecnica Aeroespacial (INTA), Ctra. de Ajalvir, km. 4, E-28850 Torrejón de Ardoz, Spain}\affiliation{Spanish Space Solar Physics Consortium}\email{alvareza@inta.es}
\author[sname='Grauf']{Bianca~Grauf} \affiliation{Max-Planck-Institut für Sonnensystemforschung, Justus-von-Liebig-Weg 3, 37077 Göttingen, Germany}\email{grauf@mps.mpg.de}
\author[sname='Carpenter']{Michael~Carpenter} \affiliation{Johns Hopkins University Applied Physics Laboratory, 11100 Johns Hopkins Road, Laurel, Maryland, USA}\email{michael.carpenter@jhuapl.edu}
\author[sname='Bell']{Alexander~Bell} \affiliation{Institut für Sonnenphysik (KIS), Georges-Köhler-Allee 401a, 79110 Freiburg, Germany}\email{albe@leibniz-kis.de}
\author[orcid=0000-0001-7764-6895,sname='Martínez~Pillet']{Valentín~Martínez~Pillet} \affiliation{Instituto de Astrofísica de Canarias, Vía Láctea, s/n, E-38205 La Laguna, Spain}\affiliation{Spanish Space Solar Physics Consortium}\email{vmpillet@iac.es}
%\author[orcid=0000-0001-7696-8665,sname='Gizon']{Laurent~Gizon} \affiliation{Max-Planck-Institut für Sonnensystemforschung, Justus-von-Liebig-Weg 3, 37077 Göttingen, Germany}\affiliation{Institut für Astrophysik und Geophysik, Georg-August-Universität Göttingen, 37077 Gōttingen, Germany}\email[show]{gizon@mps.mpg.de}

%%%% others (manually added)
\author[orcid=0000-0002-7711-5397,sname='Jafarzadeh']{Shahin~Jafarzadeh}\affiliation{Astrophysics Research Centre, School of Mathematics and Physics, Queen’s University Belfast, Belfast, BT7 1NN, UK}\email{shahin.jafarzadeh@qub.ac.uk}\affiliation{Max-Planck-Institut für Sonnensystemforschung, Justus-von-Liebig-Weg 3, 37077 Göttingen, Germany}
\author[orcid=0000-0002-9270-6785,sname='Chitta']{Lakshmi~Pradeep~Chitta}\affiliation{Max-Planck-Institut für Sonnensystemforschung, Justus-von-Liebig-Weg 3, 37077 Göttingen, Germany}\email{chitta@mps.mpg.de}

% %%% Early-Career Scientists
\author[orcid=0000-0002-7318-3536,sname='Bailén']{Francisco~Javier~Bailén} \affiliation{Instituto de Astrofísica de Andalucía, CSIC, Glorieta de la Astronomía s/n, 18008 Granada, Spain}\affiliation{Spanish Space Solar Physics Consortium}\email{fbailen@iaa.es}
\author[orcid=0000-0002-2055-441X,sname='Blanco~Rodríguez']{Julian~Blanco~Rodríguez} \affiliation{Universitat de Valencia Catedrático José Beltrán 2, E-46980 Paterna-Valencia, Spain}\affiliation{Spanish Space Solar Physics Consortium}\email{julian.blanco@uv.es}
\author[orcid=0000-0003-4319-2009,sname='Castellanos~Durán']{Juan~Sebastián~Castellanos~Durán} \affiliation{Max-Planck-Institut für Sonnensystemforschung, Justus-von-Liebig-Weg 3, 37077 Göttingen, Germany}\email{castellanos@mps.mpg.de}
\author[orcid=0009-0002-6808-5154,sname='Harnes']{Edvarda~Harnes} \affiliation{Max-Planck-Institut für Sonnensystemforschung, Justus-von-Liebig-Weg 3, 37077 Göttingen, Germany}\email{harnes@mps.mpg.de}
\author[orcid=0000-0001-6029-7529,sname='Hölken']{Johannes~Hölken} \affiliation{Max-Planck-Institut für Sonnensystemforschung, Justus-von-Liebig-Weg 3, 37077 Göttingen, Germany}\email{hoelken@mps.mpg.de}
\author[orcid=0000-0002-4669-5376,sname='Ishikawa']{Ryohtaroh~T.~Ishikawa} \affiliation{National Institute for Fusion Science, 322-6 Oroshi-cho, Toki City 509-5292, Japan}\email{ishikawa.ryohtaro@nifs.ac.jp}
\author[orcid=0000-0001-7452-0656,sname='Kawabata']{Yusuke~Kawabata} \affiliation{National Astronomical Observatory of Japan, 2-21-1 Osawa, Mitaka, Tokyo 181-8588, Japan}\email{kawabata.yusuke@nao.ac.jp}
\author[orcid=0000-0002-1043-9944,sname='Matsumoto']{Takuma~Matsumoto} \affiliation{Centre for Integrated Data Science, Institute for Space-Earth Environmental Research, Nagoya University, Furocho, Chikusa-ku, Nagoya, Aichi 464-8601, Japan}\email{takuma.matsumoto@gmail.com}
\author[orcid=0000-0002-7044-6281,sname='Oba']{Takayoshi~Oba} \affiliation{Advanced Research Center for Space Science and Technology, Institute of Science and Engineering, Kanazawa University, Kakuma-machi, Kanazawa, Ishikawa 920-1192, Japan}\affiliation{Max-Planck-Institut für Sonnensystemforschung, Justus-von-Liebig-Weg 3, 37077 Göttingen, Germany}\email{oba@mps.mpg.de}
\author[orcid=0000-0003-1483-4535,sname='Strecker']{Hanna~Strecker} \affiliation{Instituto de Astrofísica de Andalucía, CSIC, Glorieta de la Astronomía s/n, 18008 Granada, Spain}\affiliation{Spanish Space Solar Physics Consortium}\email{streckerh@iaa.es}
\author[orcid=0000-0003-1971-5551,sname='Vukadinović']{Dušan~Vukadinović} \affiliation{Institut für Physik, Universität Graz, Universitätsplatz 5, 8010 Graz, Austria}\affiliation{Max-Planck-Institut für Sonnensystemforschung, Justus-von-Liebig-Weg 3, 37077 Göttingen, Germany}\email{dusan.vukadinovic@uni-graz.at}

\begin{abstract}
	We report on the measurement of the height-dependent time shifts of wave pulses in the lower solar atmosphere from high-resolution spectro-polarimetric observations obtained with the \acs{susi} instrument on board the \sunriseiii{} balloon-borne solar observatory during its successful science flight in July 2024. The \acl{los} velocities derived from the line-core positions of \nlines{} spectral lines in a 2\,nm-wide window around the \caiih{} line were used to determine the time shifts of propagating pulses at their respective formation heights. Our analysis reveals that these shifts are roughly ordered according to the computed formation heights of the respective spectral lines.
	A statistical analysis of the time shifts using sit-and-stare observations with a total duration of one hour reveals that wave pulses propagating upwards from near the solar surface to heights of approximately 500--700\,km are most common, with average time lags of 20\,s to 30\,s between these heights. Also present are pulses with close-to-zero phase shifts, predominantly above intergranular lanes and areas of enhanced magnetic activity.
    Additionally, downward propagating wave pulses with negative time lags of 10\,s to 15\,s are seen, mostly above areas of enhanced magnetic activity. A common feature of all the observed pulses is that in the lower 250\,km they show small time lags of zero to a few seconds, and only at higher layers do the propagating pulses become more dominant. 
	This study	demonstrates the potential of the many-line approach for investigating the height dependence of the physical conditions in the solar atmosphere.

\end{abstract}

\keywords{\uat{Solar physics}{1476}}

%%%%%%%%%%%%%%%%%%%%%%%%%%%%%%%%%%%%%%%%
% This is the important file article.tex:
%!TEX root = main.tex
%%%%%%%%%%%%%%%%%%%%%%%%%%%%%%%%%%%%%%%%%%%%%%%%%%%%

\section{Introduction}
%\label{introduction}

Solar $p$-modes leave the most prominent imprints of \ac{los} oscillations in the lower solar atmosphere. They are trapped within the solar interior \cite[e.g.,][]{christensen-dalsgaard1985,christensen-dalsgaard2021}, and predominantly excited stochastically by turbulent convection in the outer layers of the Sun \cite[]{goldreich1977,goldreich1990}. As resonant pressure waves, their precise frequencies provide the fundamental data for helioseismology, allowing to probe the structure and dynamics of the Sun’s interior. 
Ground-based networks like \acsu{gong} \cite[\aclu{gong},][]{harvey1996} and space-based instruments, particularly the \ac{mdi} on the \aclu{soho} \cite[\acsu{soho},][]{scherrer1995,domingo1995}, the \ac{hmi} on the \aclu{sdo} \cite[\acsu{sdo},][]{schou12,scherrer12,pesnell:12}, and the \ac{phi} on \aclu{solo} \cite[]{solanki20,mueller20}, have extensively studied the 5-minute \ac{los} velocity oscillation pattern. These measurements primarily utilized a single spectral line, spanning a relatively narrow height range between 100\,km and 200\,km above the solar surface\footnote{For the purposes of this paper, the solar surface is defined at the layer where optical depth $\tau$ at 500\,nm reaches unity, providing only limited information on the apparent propagation of the `pulses' of the $p$- or $f$-mode waves through the solar atmosphere.}.

Below the so-called cut-off frequency and under the assumption that the perturbations are adiabatic and the atmosphere is in hydrostatic equilibrium with no motions, $p$-mode waves are evanescent, decaying with height above the surface rather than propagating vertically.
The 5-minute $p$-modes lie below the cut-off frequency in the quiet Sun, and are therefore evanescent in the photosphere \cite[]{felipe:20}.
If the perturbations in the atmosphere are adiabatic, and the atmosphere is static and uniform in the non-radial directions, then the expectation is for there to be zero phase difference between measurements of the oscillation pattern at different heights. 
In the case of the Sun, there is radiative heating and cooling, other non-adiabatic processes associated with heating and shocks, and the atmosphere contains strong horizontal gradients of velocities, temperatures, and densities.
All of these affect the evolution of acoustic modes above the surface, and non-zero phase shifts with height are both expected \cite[]{baldner2012} and have previously been observed \cite[]{zhao2012}.  Magnetic fields
are also known to play an important role \cite[]{mcintosh2001, jess2023, cally2024}.
The phase shifts contribute to what is known as the center-to-limb effect \cite[]{Christensen-DalsgaardDappen1988} and are a topic of active research \cite[]{SchouBirch2020,kitiashvili2026}.

The measurement of these time shifts is challenging, since the Dopplergrams obtained from the line center and the wings of the \acs{hmi}-\fei{} line are separated by only 60\,km to 80\,km in height \cite[]{nagashima2014}.
\cite{zhao2012} reported on a 2-second travel time difference for measurements taken at disk center and at the limb, speculating that the time lag is related to the formation height of the \fei{} 617.3\,nm line. 
\cite{baldner2012} and \cite{schou2015} attributed this center-to-limb variation to the presence of large asymmetries in up- and downflows in the solar atmosphere, causing phase shifts on the modes with height, and therefore  also with heliocentric angle.
With high spatial resolution measurements using the \aclu{ibis} \cite[\acsu{ibis},][]{cavallini2006} at the \ac{dst}, \cite{zhao2022} found that at a critical frequency of 3\,mHz (5-minute period) the time shift moves from a 1\,s lead to a 1\,s lag for waves below and above this frequency, respectively. 
\cite{waidele2023} investigated these time shifts using \ac{mhd} simulations, confirming the non-evanescence of the 5-minute $p$-modes in the photosphere and the resulting time lag. Similarly to \cite{zhao2022}, they found substantial differences between the Doppler velocities and the \ac{mhd}-plasma velocities, a consequence of the non-adiabaticity of the solar atmosphere, where a perturbation of one parameter is not instantaneously modifying other properties of the atmosphere. 

The aim of this paper is to demonstrate the potential of the high spatial and temporal resolution, multi-line observations of the \acf{susi} on \sunriseiii{} to investigate the height dependence of the physical conditions in the solar atmosphere. Here we use this unique capability for a detailed investigation of the time shifts of events propagating through the solar atmosphere, like the \ac{los} pulses of $p$- or $f$-mode waves, including the influence of magnetic fields. We measure the time lags or zero-phase shifts from the Doppler velocities obtained from the line-core positions, and attribute these to a certain height in the solar atmosphere, corresponding to the formation height of the line. The \ac{susi} observations are described in \sref{observation}, followed by the description of the analysis method in \sref{method}. Results are presented in \sref{results} with emphasis on the measured time lag and its dependence on the magnetic background. The paper concludes with a discussion of the results and a summary in \sref{discussion}. This paper is part of the early-science series of \sunriseiii{} \cite[]{solankietal2026}.

%%%%%%%%%%%%%%%%%%%%%%%%%%%%%%%%%%%%%%%%%%%%%%%%%%%%
\section{Observation}\label{observation}

We analyze a high-resolution dataset obtained with the \ac{susi} instrument \cite[]{susi25} on board the \sunriseiii{} observatory \cite[]{sunrise25}. The sit-and-stare observation was taken on 2024-07-11 between 06:52:16 and 07:51:41\,UTC (\sunriseiii{} data ID \verb|04_QSUN.SU_01a|), targeting a quiet-sun region, located close to disk center (slit center at the beginning of the time series at $x = -15$\arcsec{}, $y = +30$\arcsec{} in helioprojective coordinates), and containing some network magnetic fields. 

\figref{TuMagSUSI.pdf} shows the continuum intensity ($I_c$) map (left panel, green colormap) and Stokes-$V$ polarization map  (right panel) at -68\,m\AA{} from the nominal line core position of the \fei{} 525.02\,nm line, obtained with the \ac{tumag} instrument \cite[]{tumag25} on board the \sunriseiii{} observatory, at the start of the sit-and-stare observation. The inset with the blue colormap shows the \ac{susi} \ac{sjc} image recorded at the same time. The three vertical lines indicate the positions of the three sit-and-stare observations of the \ac{susi} instrument before the first pointing adjustment at 07:14:05~UTC (yellow, line~1), after the first pointing adjustment (cyan, line~2), and after the second pointing adjustment at 07:28:45~UTC (magenta, line~3).

\colfigwide{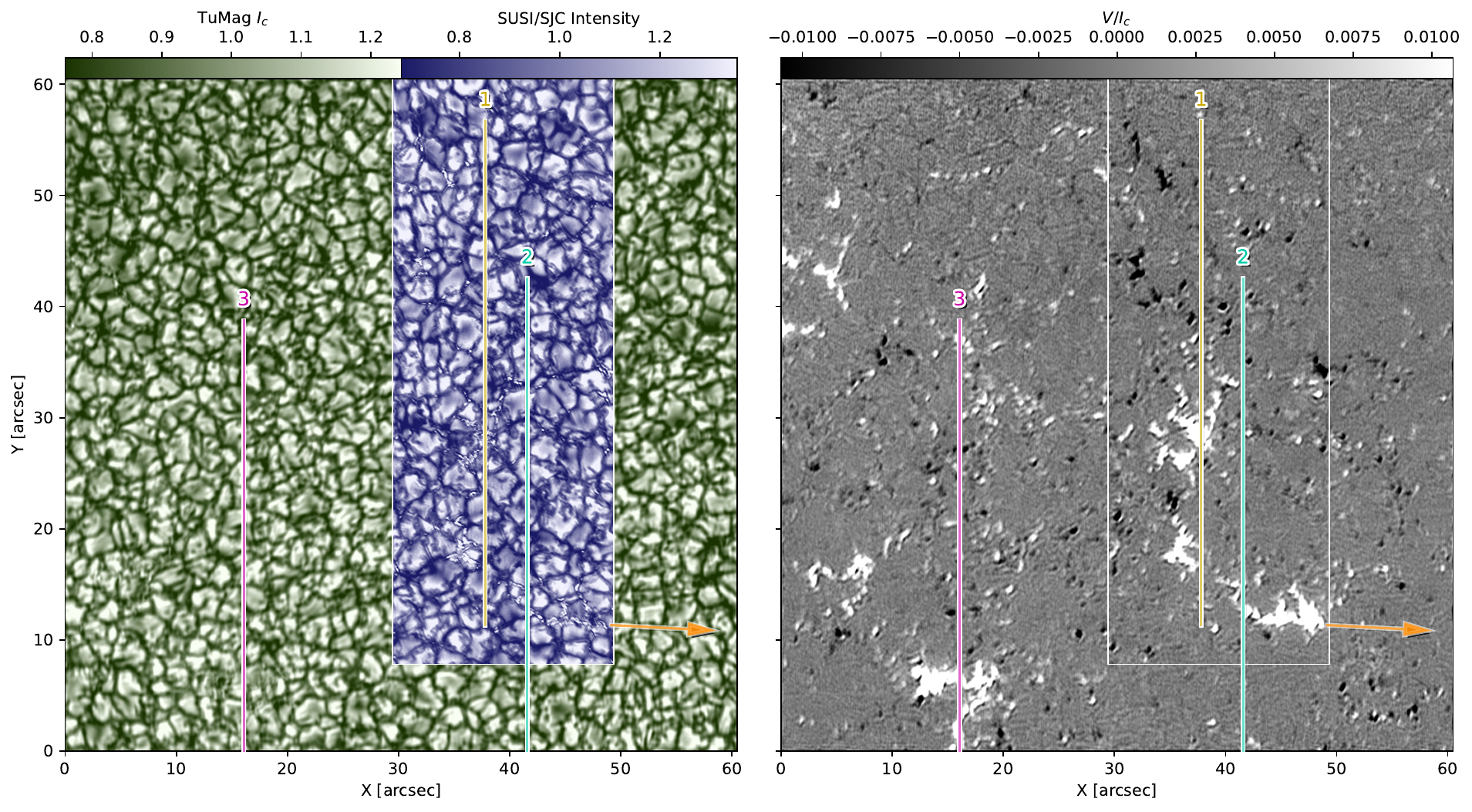}{\ac{tumag} continuum intensity (left panel, green colormap), \ac{susi} \acf{sjc} image (left panel, blue colormap) and \ac{tumag} Stokes-$V$ map (right) observed at 2024-0711T06:53:35\,UTC. The positions of the three sit-and-stare observations of \ac{susi} are indicated by the yellow (position 1), cyan (2) and magenta (3) lines. The orange arrow points towards the disk center, which is close to the tip of the arrow just a few arcseconds outside the \ac{fov}. The white rectangle indicates the \ac{susi} \ac{sjc} \ac{fov} for sit-and-stare position 1.}

The data were reduced using the standard pipelines for the \ac{tumag} and \ac{susi} instruments (see data reduction chapters in \cite{solankietal2026}, and \cite{iglesias2025}). The \ac{tumag} maps were processed with the reconstruction method described in \cite{bailen2022a,bailen2022b}.
The \ac{susi} \ac{sjc} image was deconvolved using the \aclu{momfbd} method \cite[\acsu{momfbd},][]{vanNoort2005}.
Contrast and resolution of the \ac{susi} \ac{sjc} image are naturally higher than for the \ac{tumag} maps, owing to the short wavelength of 325\,nm at which the \ac{susi} \ac{sjc} operates.

The wavelength window of this data set covered a range from 395.82\,nm to 397.77\,nm, including the \caiih{} line at 396.85\,nm. Spectral sampling was 1.01\,pm per pixel. The spatial sampling along the slit of 7\,$\mu$m width (0.06\arcsec{} on the sky) was 0.03\arcsec{} per pixel, with a slit length of 48\arcsec{}. For the analysis, the original temporal cadence of the camera of 48~\ac{fps} was reduced to the duration of two full spectropolarimetric modulation cycles to obtain a temporal resolution of $\approx$\tres{}\,s, resulting in a total of \nsteps{} time steps. This increased the photon statistics and avoided possible artifacts from the beam wobble introduced by the rotating waveplate. 

\figref{Stokes_IV.pdf} shows the observed \ac{susi} Stokes-$I$ continuum time-distance map, revealing the temporal evolution of the granules along the slit. The discontinuities correspond to the two previously mentioned pointing adjustments during the 1-hour long observation, caused by a loss of lockpoint by the \ac{cws} system \cite[]{cws26}. 

\colfig{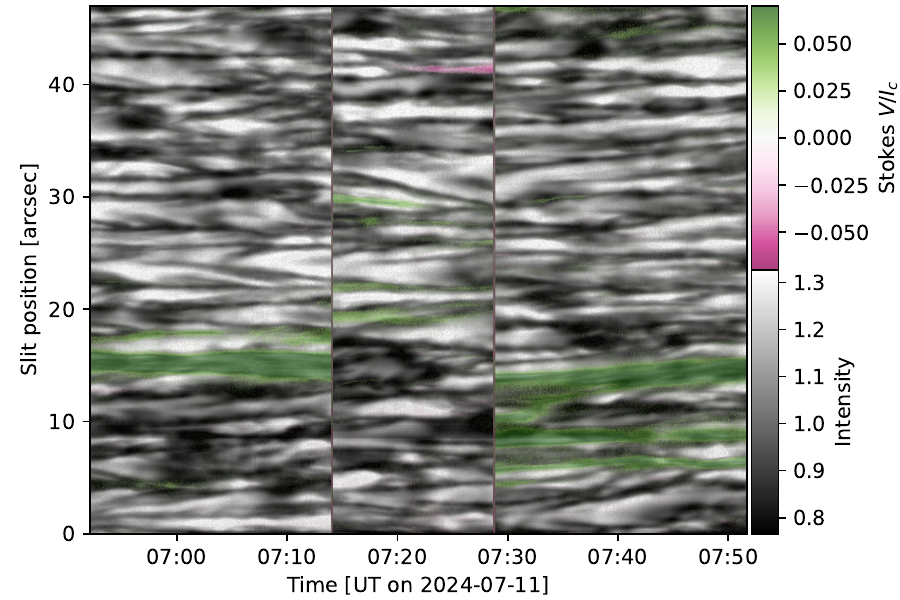}{Observed \ac{susi} continuum intensity ($I_c$) time-distance map (grayscale). The sit-and-stare observation reveals the temporal evolution of the granules. The discontinuities correspond to pointing adjustments during the 1-hour long observation. The green-purple colored overlay shows the Stokes $V/I_c$ polarization map, accumulated over all spectral lines used in the study.}

The colored overlay in \figref{Stokes_IV.pdf} shows the observed \ac{susi} Stokes-$V$ polarization map, accumulated over all spectral lines used in the study, revealing the magnetic activity along the slit. The green overlay dominates since the network regions in this quiet-sun area are mostly positive-polarity.
The regions with network fields are also visible in the $I_c$ map as an area where the granular pattern is slightly different, with smaller-scale structures, as expected for magnetically more active regions.
The Stokes-$V/I_c$ overlay was produced using single-beam demodulation with ad-hoc crosstalk correction. Wavelength calibration was performed by comparing spatially-averaged disk-center \susi{} data with the Hamburg \ac{fts} atlas \cite[]{Neckel1999} \cite[see][for details]{Hoelken2024}.

%%%%%%%%%%%%%%%%%%%%%%%%%%%%%%%%%%%%%%%%%%%%%%%%%%%%
\section{Method}\label{method}

\subsection{Spectral line selection}\label{lineselection}

The spectral region observed by \ac{susi} contains many spectral lines formed at different heights in the solar atmosphere. An average intensity spectrum is shown in \figref{height_and_peaks.pdf} (blue line), with the  \caiih{} line at 396.85\,nm in the center. From the about 100 spectral lines with prominent absorption features in the wavelength range, we hand-selected \nlines{} spectral lines that are sufficiently deep and isolated to allow the detection of oscillation signals in their line cores. The selected lines, along with their wavelengths, elements, and formation heights in a \acs{falc} atmosphere \cite[]{fontenla93} and  \acsu{muram} atmospheres \cite[\aclu{muram},][]{voegler2005,rempel14}, are listed in \tabref{tab:lines}.

\subsection{Formation heights}\label{formationheights}

The formation height versus wavelength curve presented in \figref{height_and_peaks.pdf} (orange line) shows the height where the optical depth at a given wavelength is unity in a \acs{falc} atmosphere \cite[]{fontenla93}. Under this assumption, the formation heights of the line cores of the \nlines{} selected spectral lines range from about 50\,km to 660\,km in the solar atmosphere (see column '\acs{falc}' in \tabref{tab:lines}).
The highest formation height is found for the \caiih{} line core at $\approx$1650\,km. This line, however, was not used for the analysis, since it is dominated by the 3-minute chromospheric oscillations, hiding the signal of the waves from the deeper layers. 

A more realistic formation height estimate for the selected spectral lines can be obtained from \ac{mhd} atmospheric models. We use a \acs{muram} \ac{mhd} simulation of the quiet Sun, based on a small-scale-dynamo simulation. Following \cite{rempel14,bhatia2022}, initially a weak, zero net-flux field is added. The simulation is run until the magnetic field in the box, enhanced by convective motions, saturates. An additional vertical field of 1\,G was then added, representing a weak internetwork flux imbalance.
The individual spatial pixels of this simulation contain the height-dependent atmospheric conditions for granules, intergranular lanes and regions with magnetic network, similarly to the ones observed with \sunriseiii{} / \ac{susi}.  
The formation heights of the selected spectral lines were taken  as the $\tau=1$ layer at the line core, with the spectral lines computed using the RH1.5D radiative transfer code \cite[]{uitenbroek01, pereira2015} under \ac{lte} conditions. Column '\acs{mhd}' in \tabref{tab:lines} shows the average formation height of the selected lines in the \ac{mhd} cube, with the errors representing the 25\% and 75\% percentile of the height distribution for the individual columns in the \ac{mhd} cube.

The computed line-core formation heights are used for sorting the spectral lines. In the \acs{falc} atmosphere, this sorting would be the same for the entire \ac{fov} of the \ac{susi} observation. The \ac{mhd} atmospheres, however, provide the opportunity to pick the column in the cube with similar continuum intensity level, and which best sorts the selected spectral lines, i.e., where we obtain a monotonous height dependence of the time lags. Compared to the \acs{falc} atmospheres, the formation height range expands in both directions from $\approx$15\,km to 800\,km for the \ac{mhd} atmospheres.

\colfig{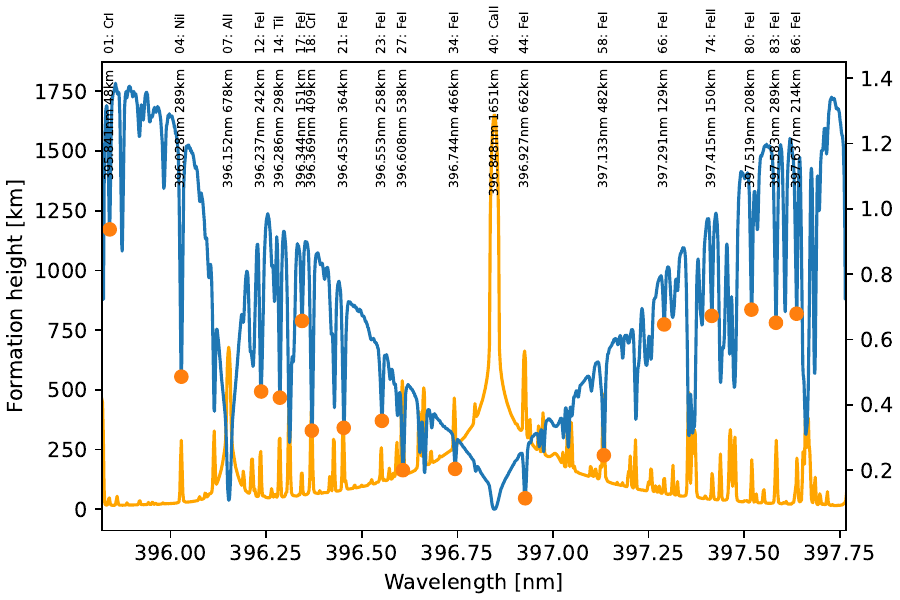}{Formation heights of the selected spectral lines in a \acs{falc} atmosphere. Blue: observed \ac{susi} spectrum, orange: formation height. The orange dots indicate the cores of the spectral lines used in this study.}

% insert table with lines, wavelengths, elements, log(gf), chi, formation heights (FALC and \ac{mhd} cube)
\begin{table}
    \centering
% Auto-generated file: git: 8ba3b91, 2026-01-29 17:35
\begin{tabular}{r|crrr}\hline
\# & Element & $\lambda$ [nm] & \multicolumn{2}{c}{Height [km]} \\
\mbox{}  & & & \acs{falc} & \acs{mhd} \\ \hline
01 & \ion{Cr}{I} & 395.839 & 47.6 & 14.6{\raisebox{0.5ex}{\tiny$^{+26.6}_{-35.9}$}} \\
74 & \ion{Fe}{II} & 397.417 & 150.1 & 93.3{\raisebox{0.5ex}{\tiny$^{+23.4}_{-21.2}$}} \\
66 & \ion{Fe}{I} & 397.291 & 128.7 & 105.3{\raisebox{0.5ex}{\tiny$^{+29.4}_{-29.5}$}} \\
17 & \ion{Fe}{I} & 396.343 & 151.1 & 125.8{\raisebox{0.5ex}{\tiny$^{+30.7}_{-31.2}$}} \\
86 & \ion{Fe}{I} & 397.639 & 214.3 & 170.5{\raisebox{0.5ex}{\tiny$^{+38.9}_{-37.8}$}} \\
80 & \ion{Fe}{I} & 397.520 & 207.7 & 182.8{\raisebox{0.5ex}{\tiny$^{+48.8}_{-47.2}$}} \\
12 & \ion{Fe}{I} & 396.235 & 242.3 & 207.5{\raisebox{0.5ex}{\tiny$^{+34.0}_{-31.7}$}} \\
04 & \ion{Ni}{I} & 396.030 & 288.8 & 231.1{\raisebox{0.5ex}{\tiny$^{+30.0}_{-31.0}$}} \\
03 & \ion{Fe}{I} & 396.028 & 288.8 & 231.1{\raisebox{0.5ex}{\tiny$^{+30.0}_{-31.0}$}} \\
83 & \ion{Fe}{I} & 397.582 & 288.6 & 234.6{\raisebox{0.5ex}{\tiny$^{+35.3}_{-32.7}$}} \\
84 & \ion{Fe}{I} & 397.584 & 288.6 & 234.6{\raisebox{0.5ex}{\tiny$^{+35.3}_{-32.7}$}} \\
23 & \ion{Fe}{I} & 396.551 & 257.7 & 216.3{\raisebox{0.5ex}{\tiny$^{+36.6}_{-31.2}$}} \\
21 & \ion{Fe}{I} & 396.452 & 363.9 & 322.2{\raisebox{0.5ex}{\tiny$^{+36.5}_{-33.8}$}} \\
14 & \ion{Ti}{I} & 396.285 & 298.0 & 424.3{\raisebox{0.5ex}{\tiny$^{+110.3}_{-127.6}$}} \\
18 & \ion{Cr}{I} & 396.368 & 409.3 & 401.9{\raisebox{0.5ex}{\tiny$^{+49.7}_{-50.3}$}} \\
34 & \ion{Fe}{I} & 396.742 & 466.0 & 435.6{\raisebox{0.5ex}{\tiny$^{+35.1}_{-36.3}$}} \\
58 & \ion{Fe}{I} & 397.132 & 482.2 & 468.0{\raisebox{0.5ex}{\tiny$^{+46.7}_{-44.4}$}} \\
27 & \ion{Fe}{I} & 396.606 & 538.1 & 583.2{\raisebox{0.5ex}{\tiny$^{+68.4}_{-73.4}$}} \\
44 & \ion{Fe}{I} & 396.926 & 662.4 & 797.1{\raisebox{0.5ex}{\tiny$^{+92.8}_{-106.8}$}} \\
\hline
\end{tabular}
% End auto-generated file
\caption{Spectral lines used for this work. The line parameters were taken from the Kurucz line list\footnote{Available online: \href{http://kurucz.harvard.edu/linelists/gfnew/}{http://kurucz.harvard.edu/linelists/gfnew/}} \cite[]{kurucz:18}, the formation height was computed for the line core in a \acs{falc} and \ac{mhd} cube atmospheres. 
\label{tab:lines}}
\end{table}

\subsubsection{Line-core fitting}\label{lincorefit}

The temporal variation of the line-core positions provide information about the velocity fluctuations of the highest layer of the solar atmosphere sampled by the corresponding spectral line. Sub-pixel wavelength accuracy is needed to determine small velocity fluctuations, requiring a reliable fitting method for the typically rather flat spectrum at the local minimum defining the line core.

We determined the line-core positions of \nlines{} selected spectral lines for the \nsteps{} time steps of the 1-hour long sit and stare observation, and for all pixels along the \ac{susi} slit. To increase the \ac{s2n} ratio, data were  accumulated to \tres{}\,s temporal resolution, and along the slit direction an interlaced binning of 4 pixels with a step size of 2 pixels was applied.
The line core was fitted in a two-step process: First, the local minimum was searched around the nominal position in the Stokes $I$ spectrum with wavelength-pixel accuracy. Then, a 6$^{\mbox{th}}$ order polynomial was fitted to $\pm 7$ wavelength points around this sampled minimum position of the spectrum, weighted with the reciprocal of the distance (plus an offset of 0.1 to avoid a division by zero) to this minimum position. The coefficients of the polynomial were used to obtain the line-core position and the corresponding line-core intensity, taken from the local minimum closest to the sampled line core position. The high polynomial order was chosen to take into account asymmetries in the line profile close to the core, ensuring the accurate determination of the line-core position. This is important to obtain the \ac{los} velocity of the highest layer of the solar atmosphere sampled by the corresponding spectral line. 

The line-core fitting results in a 2-dimensional slit-time map of line-core positions for each spectral line, which is the basis for the analysis of the wave pulses in this study. An example of such a slit-time map is shown in \figref{peakmovie_dots_StokesV.pdf} for the deepest forming spectral line (\ion{Cr}{I} 395.839\,nm, top panel). The blue-red color coding marks the amplitude of the line-core oscillation, while the green and purple overlay indicates the Stokes $V$ signal of positive and negative polarity, respectively. The same maps for all other spectral lines are shown in an \href{https://mps.files.academiccloud.de/s/ijM5Ze5HmySfQ96}{animation in the online material}. For better visibility of the oscillations, the maps are scaled individually to the 98-percentile range of each spectral line. The absolute value of the \ac{los} velocity amplitude is displayed in the colorbar of every map, with values in the range of about $\pm1.5$\,\kms{} for all spectral lines.

Horizontal cuts through the slit-time maps in \figref{peakmovie_dots_StokesV.pdf}  let the solar oscillations appear like heart-beat pulses. The middle panel of  this figure exemplifies this with a horizontal cut at $y=39$\arcsec{}, the position indicated marked with the dashed line and through the purple star symbol labeled '3'. The analysis in this study concentrates on the detection and tracking of these `heart-beat pulses'. We defined the position of the maximum redshift (the location of one maximum in the middle/bottom panel of \figref{peakmovie_dots_StokesV.pdf}) in this heart-beat line as a `pulse', since at this location the phase of the oscillations in the various spectral lines was best visible, and the time lag between the different heights could be determined most reliably. Attempts to use \ac{fft}- or wavelet-based analyses, as used in other studies \cite[e.g.,][also part of this focus issue]{Jafarzadeh2026_paper1,Jafarzadeh2026_paper2,Jafarzadeh2026_paper3}, were not successful, since the length of the three sit-and-stare observations did not cover enough wave periods. This limits also our ability to perform a frequency-dependent analysis of the time lags. Longer duration \ac{susi} datasets do exist as narrow, repetitive scans, but their temporal resolution in the range of 5\,s to 10\,s is too low for this study.

The pulse locations are marked by ($N=$\npeaks{}) dots, which are given one of three different colors according to their magnetic background. Green dots indicate oscillation above non-magnetic regions (granules), purple dots for magnetic network regions, and yellow dots for intermediate magnetic background regions. Three examples, labeled 1, 2 and 3, are shown in the bottom panel of \figref{peakmovie_dots_StokesV.pdf} as line plots, indicating the three different time-lag categories described in \sref{tlagcat}.

\colfig{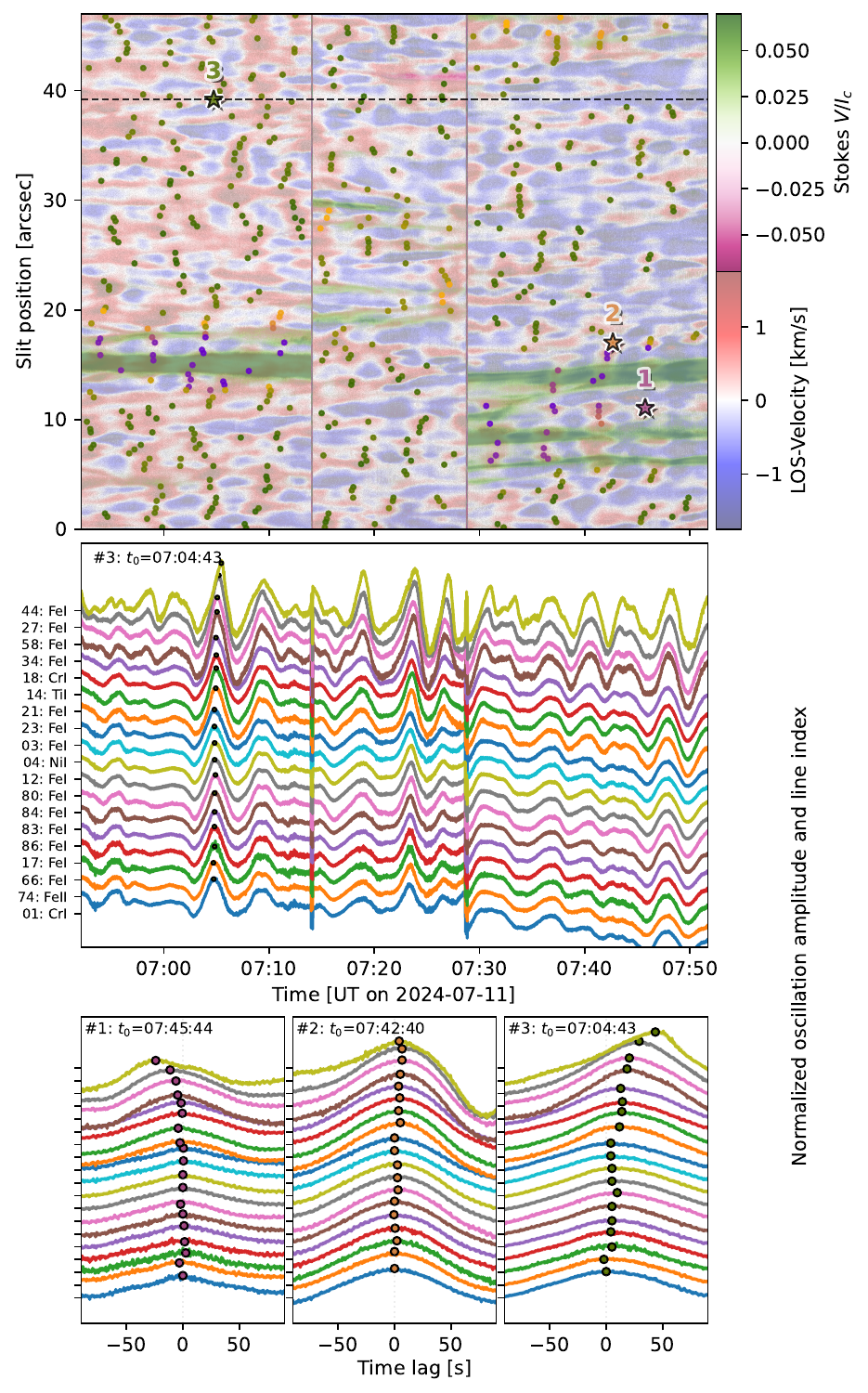}{Line-core oscillation maps and line plots. Top panel: Slit-time map for the deepest forming spectral line (\ion{Cr}{I} 395.839\,nm). The blue-red color coding marks the amplitude of the line-core \ac{los}-velocity oscillation. The green and purple overlay indicates the Stokes $V$ signal of positive and negative polarity, respectively. The velocity peak locations are marked by dots ($N=$\npeaks{}). An animation showing the maps for all other spectral lines is available online at \href{https://mps.files.academiccloud.de/s/ijM5Ze5HmySfQ96}{this link}. The animation starts $h=$14.6\,km and ends at $h=$797.1\,km. At the top of then animation, the line number, line identifier, and $h$ value is provided. The real-time duration of the animation is 6 seconds.
Middle panel: Horizontal cut through top panel (dashed line), containing the location (3) at ($y$=39\arcsec{}, $t_0$=07:04:43), showing the line-core oscillations for all \nlines{} spectral lines.
Bottom panel: Zooms into locations (1), (2), and (3), labeled in the top panel, and exemplifying the three $t$-lag categories for the propagation of wave pulses in the solar atmosphere: downward propagation (left, $t$-lag category 1), zero propagation (middle, $t$-lag category 2) and upward propagation (right, $t$-lag category 3).}

%%%%%%%%%%%%%%%%%%%%%%%%%%%%%%%%%%%%%%%%%%%%%%%%%%%%
\section{Results}\label{results}

The oscillation maps for all spectral lines, exemplified in the top panel of \figref{peakmovie_dots_StokesV.pdf} for the deepest forming line (\ion{Cr}{I} 395.839\,nm), reveal the presence of wave pulses in the solar atmosphere. The regular occurrences of the peaks are consistent with the 5-minute oscillations on the solar surface, the dominant period of $p$-modes in the quiet Sun. The accurate determination of the line-core positions allows us to track the apparent propagation of the individual wave pulses through the different layers of the solar atmosphere, as sampled by the selected spectral lines. This reveals the time lag of the pulses as they propagate through the atmosphere.

The height nodes for determining the time lag were defined by the formation heights of the spectral lines, computed from the \ac{mhd} cube. The best matching \ac{mhd} column was selected for each pulse, imposing the following  criteria: (1) The \ac{mhd} column is required to be in the same range of continuum intensity values as the observed pulse locations ($\pm0.07$ of the observed continuum intensity). This ensures that the atmospheric conditions in the simulation are similar to those in the observations. (2) The \ac{mhd} column is required to lie in a region of low spatial variation in the \ac{mhd} cube. This ensures that the selected \ac{mhd} column is not affected by horizontal gradients, potentially altering the height stratification. We also tried to use Stokes $V/I_c$ as the selection criterion, but found this to sort the data worse. Obviously, the influence of the weak magnetic fields in both, the simulation and the observed region, on the atmospheric structure is significantly smaller than that of the temperature in the continuum forming layer.

\subsection{Time-lag categories}\label{tlagcat}

As mentioned in the introduction, time shifts of the wave pulses at different heights are expected due to multiple reasons: the atmosphere is dynamic, and non-adiabatic processes such as radiative transport are important. It is convenient to classify the different types of height dependence we see, and for this reason we will use the terms downward propagation, zero propagation, and upward propagation based on how the phase is shifted from line to line based on its formation height. This is intended as a shorthand, the structure of the modes in the presence of e.g. shocks, radiative transfer, and magnetic fields is complicated and remains a topic of active research \cite[]{kitiashvili2026}. 

With the above caveat in mind, we find three distinct time-lag categories for the wave pulse propagation, shown in the examples in the bottom panel of \figref{peakmovie_dots_StokesV.pdf} as zoom-ins similar to the `heart-beat' lines shown in the middle panel, at three different $(y,t)$ positions. The first category (bottom left, $t$-lag category 1) clearly shows a downward propagation of the wave pulse, i.e., the pulse appears first at the highest forming spectral line, and then moves towards the lower forming spectral lines (negative time lag). 
For the second category (bottom middle, $t$-lag category 2), no time lag is observed between the peaks of the different spectral lines, revealing no clear propagation pattern (zero phase difference). 
The third category (bottom right, $t$-lag category 3, positive time lag) shows an upward propagating wave pulse, with the positive time lag of the peaks best visible in the middle to higher forming spectral lines. The $t$-lag categories were defined by setting  thresholds for the time lag of the highest forming spectral lines (500\,km to 700\,km) with respect to the deepest forming line (\ion{Cr}{I}, 0\,km to 50\,km). The threshold  was set to $\pm$\tlagthresh\,s, a value separating the three categories well ($t$-lag category 1 with time lags below -\tlagthresh\,s, $t$-lag category 2 between -\tlagthresh\,s and +\tlagthresh\,s, and $t$-lag category 3 above +\tlagthresh\,s). 

The average phase speeds computed from the time lag and the height difference between the highest and lowest forming spectral lines have a lower limit of about 10\,\kms{} for the largest time lags measured in category 3, and naturally reach extreme values when approaching the zero time lag in category 2. These speeds are significantly higher than the sound speed in the solar atmosphere, assumed to be in the range of 7\,\kms{} to 8\,\kms{} in the upper photosphere. This indicates that the measured time lags are not related to the actual propagation of a wave pulse through the solar atmosphere.

\colfigtwo{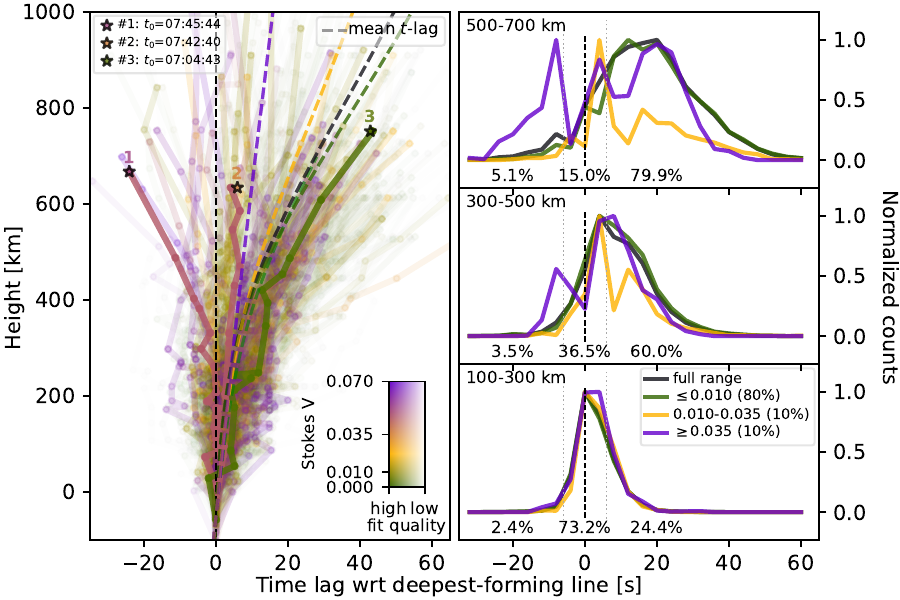}{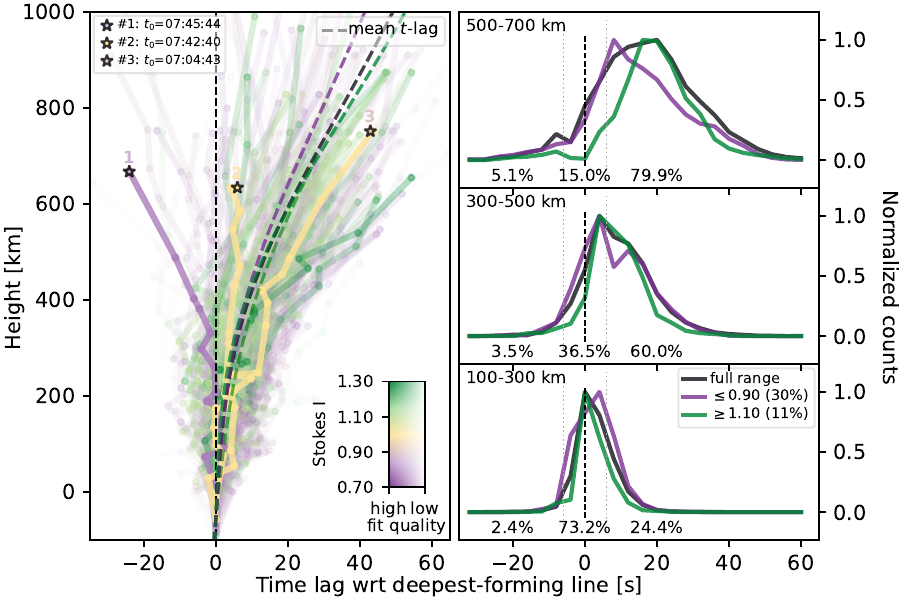}{Height dependence of time lags for all \npeaks{} traced pulses  with respect to the peak position of the deepest forming spectral line ($t_0$). Upper left panel: The individual lines connect the time lags of a single pulse measured at the line-core formation heights of all spectral line computed from the best-matching \ac{mhd} atmosphere. The three examples shown in the bottom panel of \figref{peakmovie_dots_StokesV.pdf} are highlighted with increased opacity, with their highest point marked with a star. Upper right panel: Histograms of time lags for three different height ranges. Bottom: Same as top panel,  but now with the continuum intensity as the color-shading agent. }

\figref{t-lag_MHD_StokesV.pdf} traces the time-lags of all \npeaks{} pulses. With the threshold of $\pm$6\,s mentioned above, \ncatone{} of the pulses show downward propagation, \ncattwo{} show no radial propagation, and \ncatthree{} show upward propagation. These percentages are valid for the highest layer in \figref{t-lag_MHD_StokesV.pdf} (upper right panel, histograms over heights from 500\,km to 700\,km), where the time lag of the pulses is highest and therefore best visible. The percentages for the lower layers are also specified in the histograms on the right panel of \figref{t-lag_MHD_StokesV.pdf}, left, between and right of the dotted vertical lines at $\pm$\tlagthresh{}\,s, respectively.

\subsection{Magnetic background}\label{mag_bg}

The color coding of the lines in the top panel of \figref{t-lag_MHD_StokesV.pdf} describes the magnetic activity level at the spatial locations of the pulses, determined from the underlying Stokes-$V$ signals at the start times of the pulses (see colored areas in \figref{Stokes_IV.pdf}), ignoring the Stokes-$V$ polarity. Purple color indicates a strong Stokes-$V$ signal (integrated $V/I_c\ge\vhigh$), green the very quiet internetwork regions with weak Stokes-$V$ signals ($V/I_c\le\vlow$), and yellow the regions of intermediately strong magnetic background.

The dots in the left panel of \figref{t-lag_MHD_StokesV.pdf}, connected by the colored lines, show the position of the pulses for every spectral line. On average, the position of these points can be fitted well with a cubic spline with three nodes at heights -50\,km, 375\,km, and 800\,km. The second derivative of the spline at the endpoints was set to zero, mimicking a constant propagation speed of the pulse at the deepest layer (set to -50\,km) and the highest forming spectral line, respectively. This so-called natural spline boundary condition allows for fitting the observed decrease in propagation speed with height, visible in the left panels of \figref{t-lag_MHD_StokesV.pdf}. 

The zig-zag fluctuations around the smooth spline fit can be attributed to the errors in the formation height estimates for the spectral lines, used for the sorting of the time lag with height, and to the determination of the exact time position of the pulse, which often is rather shallow and broad. The \ac{rms} fluctuation of the measured time lag with respect to the corresponding spline fit is below 6\,s for the best 25\% of the pulses. This defined the threshold used to separate the three $t$-lag categories in \sref{tlagcat}.
For better visibility, we use the residual to the spline fit as a weighting factor for the histograms on the right panel of \figref{t-lag_MHD_StokesV.pdf}, with the pulses with  smallest residuals contributing more to the histograms. Similarly, the transparency of the lines in the left panels of \figref{t-lag_MHD_StokesV.pdf} is weighted with this residual, with the pulses with the smallest residuals shown with the highest opacity. Histograms and color-coding are both normalized to the number of the pulses in each Stokes-$V$ bin, with the percentages of the three bins specified in the legend of the bottom right histogram. \vbingreen{} of the pulses fall into the bin with the weakest Stokes-$V$ signal ($V/I_c\le\vlow$), \vbinpurple{} in the strongest one ($V/I_c\ge\vhigh$), and \vbinyellow{} in between.

Upward propagating pulses ($t$-lag category 3) are present at all magnetic activity levels, with an average time-lag for the propagation from the lowest to the highest sampled layer of about 20\,s to 30\,s. With increasing magnetic activity, the time lag moves towards lower values, with a mean value of $\approx$10\,s for intermediate magnetic activity (yellow lines). Downward propagating pulses ($t$-lag category 1) are preferentially located in regions with stronger Stokes-$V$ signals (purple lines).

The downward propagating pulses are sometimes accompanied by a second peak in the line-core oscillation, corresponding to an upward propagating pulse directly following the downward propagating pulse. An indication of this double-peak pattern can be seen in the bottom left panel of \figref{peakmovie_dots_StokesV.pdf}, with the second peak becoming visible in the highest forming spectral line with a negative time lag of about 15\,s.

\subsection{Continuum intensity background}\label{icont_bg}

Similar to \sref{mag_bg}, we now investigate the dependence of the time lag with respect to the continuum intensity ($I_c$), pointing to possible differences in the propagation patterns when located above bright granules or intergranular lanes. The bottom panel of  \figref{t-lag_MHD_StokesV.pdf} shows the same results as the top panel, but now with the continuum intensity as the color-shading agent. Green indicates the brightest regions ($I_c>1.1$), magenta the intergranular lanes ($I_c<0.9$). The regions with a continuum intensity around one are shown in light yellow.

The histogram for the highest layer in the bottom panel of \figref{t-lag_MHD_StokesV.pdf} reveals that the pulses above intergranular lanes ($I_c<0.9$) show on average smaller time lags than the pulses above granules ($I_c>1.1$). This is consistent with the dependence of the time lag on the magnetic background, described in \sref{mag_bg}, since the stronger magnetic fields are typically located in the intergranular lanes.

\section{Discussion and Conclusions}\label{discussion}

The many-line observations of \sunriseiii{} allow extending the investigation of the height-dependent apparent propagation of \ac{los} velocity pulses, in particular from the pulses of $p$- or $f$-mode waves, in the lower solar atmosphere, as well as studying the dependence of the propagation patterns on the underlying atmospheric conditions, such as the magnetic field and the continuum intensity. Whereas previous studies used the bisector analysis technique in a single spectral line, we now use the line-core formation heights as sampling points within the solar atmosphere. 

The average time lag reported by \ac{hmi} is in the 1 to 2-second range over the formation height of the \fei{} 617.3\,nm line. This time lag is in the range of the helioseismic center-to-limb effect. Our study confirms this value: \figref{t-lag_MHD_StokesV.pdf} shows that over a height range from 100\,km to 270\,km, corresponding to the range covered by the \fei{} 617.3\,nm line, the measured time lag is in a similar regime. This can be read from the average time lags in the left panels of both figures, indicated by the dashed spline fits to the averaged time lags for the various background conditions. 

We also find good agreement with the results of \cite{Fleck1989}, who performed a study of the phase differences for two chromospheric \ion{Ca}{II} lines in the near-infrared and a photospheric \ion{Fe}{I} line. They also report on very high phase speeds of 30\,\kms{} for a height range between 350\,km and 800\,km. The conversion of their phase differences to our time lags yields values of 4\,s to 11\,s for frequencies from 3\,mHz to 5\,mHz, with upward propagating pulses, similar to our $t$-lag category 3. At frequencies slightly below 3\,mHz, the waves are evanescent (zero phase shift), and below 2.5\,mHz, the waves show downward propagation (positive phase shift with the definition of \cite{Fleck1989}), similar to our $t$-lag category 1. Both, the time lags and the distribution of the propagation directions, are in general agreement with our results.

Compared to the bisector analyses of previous studies, the \ac{susi} observation covers a much larger height range, from the low photosphere to $\lesssim$1000\,km. Also, the high spatial resolution of the \susi{} data allows us to distinguish the time lags of the propagating pulses according to the prevailing conditions in the lower solar atmosphere. We find that in the intergranular lanes ($I_c<0.9$) the time lags in the highest measured layers are with $\approx$10\,s on average only half of the time lags in the granules ($I_c>1.1$, $\approx$20\,s). 

Similarly, the effect of the underlying magnetic background could be investigated. An interesting finding is the negative time lag in $\approx$5\% of the pulses,  predominantly observed above stronger magnetic field regions (line-accumulated Stokes-$V/I_c$ signal above 3.5\%), and visualized with the purple propagation lines in the left half of \figref{t-lag_MHD_StokesV.pdf}. 

However, the majority of the $p$- or $f$-mode pulses (about 80\%) shows zero or slightly positive phase shifts in the lower layers ($\lesssim$250\,km), irrespective of their magnetic or intensity background, consistent with evanescent waves. 
At higher layers, the phase speeds of the waves exceed values of 10\,kms{}, higher than what is expected for typical propagating waves, such as traveling sound waves. Height-dependent phase shifts caused by asymmetries in the quiet-sun granulation, as described in \cite{baldner2012}, are one possible explanation for these apparent high speeds. Another one is that the high speeds result from the non-adiabatic nature of the solar atmosphere. The eigenfunctions of the Sun differ from those of an adiabatic system, where evanescent waves exhibit infinite propagation speed in the radial direction. The non-adiabaticity removes energy above the surface, introducing a phase lag (as expected from damping). This damping effect removes energy that must be carried up from below. In this sense, the wave is apparently ‘propagating’. However, if the non-adiabaticity is small, the waves become almost evanescent with a very high ‘propagation speed’.

The many-line observations of \ac{susi} open a new diagnostic window for wave-pulse propagation studies. The full potential of the method is not yet exploited: higher data reduction levels will not only allow for higher-spatial resolution studies, but also for a better investigation of the magnetic background, including strength and direction of the magnetic field. 
Many-line full-Stokes inversions, specifically tailored to \susi{}’s near-ultraviolet diagnostics \cite[]{riethmueller2019,hoelken2026}, will enable a more comprehensive investigation of the height-dependence of various phenomena in the lower solar atmosphere. These novel approaches hold the potential to gain novel insights into magnetic field oscillations and the vertical and horizontal structure of field strength and inclination on small spatial scales.

%%%%%%%%%%%%%%%%%%%%%%%%%%%%%%%%%%%%%%%%

%!TEX root = main.tex

\begin{acknowledgments}

We thank the Max Planck Society for the financial support. A special thanks goes to the Max-Planck-Förderstiftung, especially to Frau Pfündl, Frau Philipp and Prof. Pöllath, for their enthusiasm for \sunriseiii{} and the generous funding of this research. We thank the employees of \ac{csbf} and \ac{esrange} / \ac{ssc} for their support and hospitality during the launch campaigns. 
%We warmly thank the kind and professional help of the personnel at Calar Alto Astronomical Observatory (Spain) who re-aluminized the \sunrise{} mirrors. 
%We are grateful to C.~Müller and M.~Wölfle of \ac{imtek} (Freiburg, Germany) for excellent technical support in diamond milling of the \ac{hrw}, and Mike Smith from Aerostar International (TX, USA) for the excellent support during the 2024 campaign for providing the wind shield film for the \acp{erack} and the \ac{pfi}.

%This version of the article has been accepted for publication, after peer review but is not the Version of Record and does not reflect post-acceptance improvements, or any corrections. The Version of Record is available online at: \href{https://doi.org/10.1007/s11207-025-02485-1}{https://doi.org/10.1007/s11207-025-02485-1}.

\sunriseiii{} is supported by funding from the Max-Planck-Förderstiftung (Max Planck Foundation), NASA under Grant \#80NSSC18K0934 and \#80NSSC24M0024 (``Heliophysics Low Cost Access to Space'' program),  and the ISAS/JAXA Small Mission-of-Opportunity program and JSPS KAKENHI JP18H05234/JP23H01220. 
This research has received financial support from the European Union’s Horizon 2020 research and innovation program under grant agreement No. 824135 (SOLARNET), and Nos. 101097844 (WINSUN) and 101039844 (ORIGIN) from the European Research Council (ERC). It has also been funded by the Deutsches Zentrum für Luft- und Raumfahrt e.V. (DLR, grant no. 50 OO 1608).
The Spanish contributions have been funded by the Spanish MCIN/AEI under projects RTI2018-096886-B-C5, and PID2021-125325OB-C5, and from “Center of Excellence Severo Ochoa” awards to IAA-CSIC (SEV-2017-0709, CEX2021-001131-S), all co-funded by European REDEF funds, "A way of making Europe”.

A. Siu-Tapia acknowledges funding from the Consejer\'ia de Transformaci\'on Econ\'omica, Industria, Conocimiento y Universidades of the Junta de Andaluc\'ia through grant POSTDOC-21-00832.
D.~Orozco~Suárez acknowledges financial support from a {\em Ram\'on y Cajal} fellowship.
The contributions of J.~H\"olken, D.~Vukadinović and E.~Harnes have been supported by the International Max Planck Research School for Solar System Science at the University of Göttingen (IMPRS), Germany.      
F.\,A.~Iglesias is a member of the ``Carrera del Investigador Cient\'ifico" of CONICET and supported by \ac{mps} through the Max Planck Partner Group between \ac{mps} and the University of Mendoza, Argentina.
The research activities and the flight operation of the SCIP team members, R.~T. ~Ishikawa, M.~Kubo, Y.~Kawabata, and T.~Oba, have been supported from the JSPS KAKENHI grants No. 23KJ0299, No. 24K07105, No. 23K13152, and No. 21K13972, respectively.  L.P.C. gratefully acknowledges funding by the European Union (ERC, ORIGIN, 101039844).
S. Jafarzadeh acknowledges support from the UK Science and Technology Facilities Council (STFC) through consolidated grants ST/T00021X/1 and ST/X000923/1.
\end{acknowledgments}

\bibliography{main}{}
\bibliographystyle{aasjournalv7}

\newpage

%!TEX root = main.tex
%%%%%%%%%%%%%%%%%%%%%%%%%%%%%%%%%%%%%%%%%%%%%%%%%%%%

\appendix

\change{\textbf{Caption for animated version of \figref{peakmovie_dots_StokesV.pdf}}: 
The animation (\href{https://mps.files.academiccloud.de/s/ijM5Ze5HmySfQ96}{link}) shows the evolution of the line-core oscillation map (top panel of \figref{peakmovie_dots_StokesV.pdf}), when stepping through the different layers of the solar atmosphere sampled by the spectral line indicated in the title of each frame. The first frame shows the deepest layer (\ion{Cr}{I} 395.839\,nm, $h=15$\,km), the last frame shows the highest layer (\ion{Fe}{I} 396.926\,nm, $h=797$\,km). The red and blue background color coding marks the \ac{los}-velocity of the corresponding line-core, the purple and green overlay indicates the Stokes $V$ signal of negative and positive polarity, respectively. The dots mark the locations of the local maxima in the \ac{los}-velocity maps (the "pulses"). The color-coding for the dots is the same as in \figref{peakmovie_dots_StokesV.pdf} with green dots indicating a pulse above non-magnetic regions, purple dots for magnetic network regions, and yellow dots for intermediate magnetic background regions.
}

% \input{appendix}
% \section{Acronyms}

%Required for the acronyms in the text



\end{document}